\documentclass{acm_proc_article-sp}
\usepackage{listings}
\newcommand{\slet}{\texttt{SelfLet}}
\newcommand{\moreslet}{\texttt{SelfLet}s}

\lstdefinelanguage{drools}{
morekeywords = {when,then,begin,end,rule,function,return},
sensitive=false, 
morecomment=[l]{//},
morecomment=[s]{/*}{*/},
tabsize=4,
basicstyle=\footnotesize,
columns=[r]{fullflexible}
}

\begin{document}

\conferenceinfo{}{Permission to make digital or hard copies of all or part of this work for personal or classroom use is granted without fee provided that copies are not made or distributed for profit or commercial advantage and that copies bear this notice and the full citation on the first page. To copy otherwise, or republish, to post on servers or to redistribute to lists, requires prior specific permission and/or a fee. \\
ROSSA 2009, October 19, 2009 - Pisa, Italy.\\
Copyright 2009 ICST 978-963-9799-70-7\/00\/0004 \$5.00.
}

\title{Incorporating prediction models in the SelfLet framework: a plugin approach}

\numberofauthors{1} \author{
\alignauthor
Nicol\`{o} Maria Calcavecchia, Elisabetta Di Nitto\\
       \affaddr{Politecnico di Milano}\\
       \affaddr{Dipartimento di Elettronica e Informazione}\\
       \affaddr{Piazza Leonardo da Vinci, 32, 20133 Milano, Italy}\\
       \email{\{calcavecchia,dinitto\}@elet.polimi.it}
}

\maketitle

\begin{abstract}
A complex pervasive system is typically composed of many cooperating \emph{nodes}, running on machines with different capabilities, and pervasively distributed across the environment. These systems pose several new challenges such as the need for the nodes to manage autonomously and dynamically in order to adapt to changes detected in the environment. To address the above issue, a number of autonomic frameworks has been proposed. These usually offer either predefined self-management policies or programmatic mechanisms for creating new policies at design time. From a more theoretical perspective, some works propose the adoption of prediction models as a way to anticipate the evolution of the system and to make timely decisions. In this context, our aim is to experiment with the integration of prediction models within a specific autonomic framework in order to assess the feasibility of such integration in a setting where the characteristics of dynamicity, decentralization, and cooperation among nodes are important. We extend an existing infrastructure called \emph{SelfLets} in order to make it ready to host various prediction models that can be dynamically plugged and unplugged in the various component nodes, thus enabling a wide range of predictions to be performed. Also, we show in a simple example how the system works when adopting a specific prediction model from the literature. 
\end{abstract}




\section{Introduction}

The dramatic diffusion of portable devices and embedded sensors, together with the increasing wide availability of networked environments, enable a tight integration and collaboration among devices and between these and more powerful software systems running on normal computing units. Thus, a complex pervasive system is typically composed by many cooperating \emph{nodes}, running on machines with different capabilities, and pervasively distributed across the environment. These systems pose several new challenges such as the need for the nodes to manage autonomously and dynamically in order to adapt to changes detected in the environment as suggested by the autonomic computing field~\cite{autonomic-vision}. The self-management mechanisms that have to be put in place can be characterized by the following core properties:
\begin{description}
\item[Dynamicity] Nodes and connections can appear and disappear at runtime in an unpredictable way (node churn)~\cite{SR06}, thus resulting in the fact that self-management has to be highly dynamic.
\item[Decentralization] The large number of nodes involved makes a centralized control mechanism infeasible. Nodes must, instead, communicate in a peer-to-peer fashion and make individual decisions to avoid bottlenecks and single points of failure. 
\item[Cooperation] Given the number and different capabilities of the devices running nodes,  various self-management elements should cooperate to share and optimize the use of computational resources without the intervention of an administrator.
\end{description}

In order to address the above issues, a number of autonomic frameworks to support the development of distributed complex systems have been proposed~\cite{automate, anthill}. They usually offer either predefined self-management policies or programmatic mechanisms for creating new policies at design time. From a more theoretical perspective, some works propose the adoption of prediction models~\cite{VAHMW02} as a way to anticipate the evolution of the system and to make timely decisions. The prediction models available in the literature typically focus on specific problems to which they apply specific, often different techniques. 

In this context, our aim is to experiment with the integration of prediction models within a specific autonomic framework in order to assess the feasibility of such integration in a setting where the characteristics of dynamicity, decentralization, and cooperation among nodes are important. In particular, we base our work on the \emph{SelfLets} infrastructure that we have presented in~\cite{ASE07,ARAMIS08}. We extend such infrastructure in order to make it ready to host various prediction models that can be dynamically plugged and unplugged in the various component nodes, thus enabling a wide range of predictions to be performed. Also, we show in a simple example how the system works when adopting a specific prediction model from the literature. 

The remainder of this paper is organized as follows: Section \ref{sec:prediction_models} describes the problem of applying prediction models to autonomic systems. Section \ref{sec:selflet_approach} presents our reference autonomic framework. Section \ref{sec:framework} shows the contribution of this work to the existing system. Section \ref{sec:example} shows an example showing the capabilities of the proposed framework. Section \ref{sec:related_work} describes some relevant similar works and Section \ref{sec:conclusion} presents our conclusion and future work.
\section{Runtime prediction models for self-organizing systems}
\label{sec:prediction_models}

The Self-* properties envisioned by the autonomic computing Manifesto \cite{manifesto} represent high-level objectives that will be gradually achieved following an evolutionary approach rather than a revolutionary one \cite{ganek_corbi}. Much of the current research focuses in developing models able to describe the characteristics of systems enhanced with autonomic features. In this context, the adoption of prediction models represents a valid direction to achieve the Self-* properties; in fact, having a good estimate of the future internal state and external world allows the system to adjust the current behavior based on self-optimization and self-healing criteria.

In the literature a large number of prediction techniques ranging from time-series analysis \cite{BJR94} to machine-learning \cite{FGGPLS} are available. The choice of the most appropriate algorithm depends at least on two factors \cite{VAHMW02}: the predicted time horizon (long-term vs. short-term) and the type of data (e.g., numerical, Boolean). However our efforts do not concentrate on the development of a new algorithm. Instead, we focus on the development of a framework able to host different techniques without being strictly tied to a particular one. This approach has the advantage of being flexible and to represent a useful testbed for studying the impact of predictive algorithms on the self-* properties.

When applying prediction models to autonomic systems important points to be investigated concern the way the information used by prediction algorithms are collected and handled, where the prediction algorithm runs, and where the subsequent decisions are made. When the system is distributed, we can easily imagine that pieces of information are spread across all nodes. In this situation, it is not always feasible and efficient to convey all such pieces to a single node that runs the prediction algorithm, as it is usually assumed. On the contrary, it may be more convenient to keep information local to each node (or meaningful subset of nodes) and to perform predictions in each node. Of course, in this case, as each element has an incomplete view of the system state, the task of producing valid predictions becomes more challenging and may produce non-optimal results. 

Estimated results obtained from prediction mechanisms are then employed by policies which regulate the specific reaction; again, the decision on which reaction to apply can be made in a centralized or decentralized way.  

While most of the approaches in the literature take a centralized perspective, we implement our framework in order to support also the local perspective with the aim of comparing the two approaches and trying to identify the circumstances under which the local one produces good results.
\section{The SelfLet approach}
\label{sec:selflet_approach}

The autonomic framework that we use as a starting point for our work is presented in more details in~\cite{ASE07,ARAMIS08} and is based on an autonomic infrastructure and architectural model called \moreslet. 

According to the \moreslet\ model a software system is composed of various autonomous elements, each of which is a \slet{}. Every \slet{} is defined in terms of: 
\begin{itemize}
\item
A \emph{main goal} that defines the aim for which it has been developed. 
\item
A \emph{main behavior} that represents a possible implementation to fulfill the main goal. It is continuously executed by the component and may depend on some external services offered either by the same \slet{} or by other \moreslet{}. 
\item
A number of \emph{offered services} and the corresponding implementation (\emph{behaviors}) that can be executed in parallel with the main behavior and are triggered on demand by the component itself or by its neighbors through a service call. 
\item
Some \emph{autonomic policies} that define reactions to abnormal situations that can occur during the life-time of the \slet. The reactions include the possibility of changing the main behavior, disabling/enabling the execution of the other behaviors, disabling/enabling the interaction with some neighbors, etc. 
\end{itemize}
In our approach, services can be offered in different ways. A \textit{SelfLet} can run a Service and return the result to the caller: in this case the service is offered in a ``Can Do'' way. Alternatively, the \textit{SelfLet} can be available to \textit{teach} the service, so that the requester is able to execute the service by itself from then on. We call this the ``Can Teach'' way. A \textit{SelfLet} can offer a service even if it does not know it directly, by using the ``Knows Who Can Do'' and ``Knows Who Can Teach'' ways: in this case, the \textit{SelfLet} will give the requester information about \textit{SelfLets} able to either offer the service or provide directions for the execution of the service. Each combination of these ``offer modes'' is allowed. 

\textit{SelfLets} can respond to requests for a specific service call or spontaneously advertise their offered services to inform the other \textit{SelfLets} about them. Indeed, when issuing a service request, a \textit{SelfLet} can specify the offer mode it would prefer. The preference it expresses depends in most cases on its current policy and on the actual need for that service.

Each \textit{SelfLet} maintains a list of known providers for each service it requires (and cannot offer itself). Then, when a need for such a service arises, the \textit{SelfLet} selects one of these providers according to a given policy, and directly asks it for the service. The \textit{SelfLets} in a network cooperate to keep such lists always up-to-date with information about the availability of providers.

The \textit{SelfLet} offers to policies writers a list of actions that enable the transformation of several aspects of the \textit{SelfLet} itself. These methods are the way the decisions made by the policies are put in action to adapt to a changed situation. The actions offered in the current implementation include, besides the methods related to  service execution:
\begin{itemize}
\item
change the way a  service is offered or asked. In particular, it could be offered or asked in a \textit{teach}, \textit{do}, or \textit{know who can do} mode, or in all modes at the same time;
\item
install a new service within the \textit{SelfLet};
\item
install new abilities;
\item
modify a given behavior, directly deleting and replacing its component states and transitions.
\end{itemize}

\begin{figure}[htb]
    \begin{center}
      \includegraphics[width=\columnwidth]{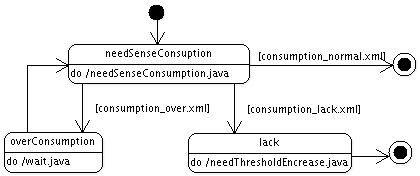}	
      \caption{\label{fig:example} Main behavior of the Electricity \slet.} 
    \end{center}
  \end{figure}

Designers program the behaviors using UML state diagrams (with some restrictions that have been introduced to formalize their semantics) and the autonomic policies using a rule language called Drools~\cite{Drools}. Figure \ref{fig:example} shows an example of a behavior taken from a \slet{} managing consumption of energy in a home environment. In the initial state the current level of energy consumption is monitored and, depending on this value, three things may happen. The conditions over the arcs define what the next state is by comparing the sensed energy value to an internal threshold. If the consumption is normal then no action is taken; if the sensed value is dangerously near to a maximum threshold then a request for more energy capacity is issued to the other neighbor \moreslet{}; finally, if the current consumption is higher than the threshold then the \slet{} goes into a wait state where the energy provisioning to the corresponding home is blocked.

\begin{figure}[htp] \centering 
\includegraphics[width=\columnwidth]{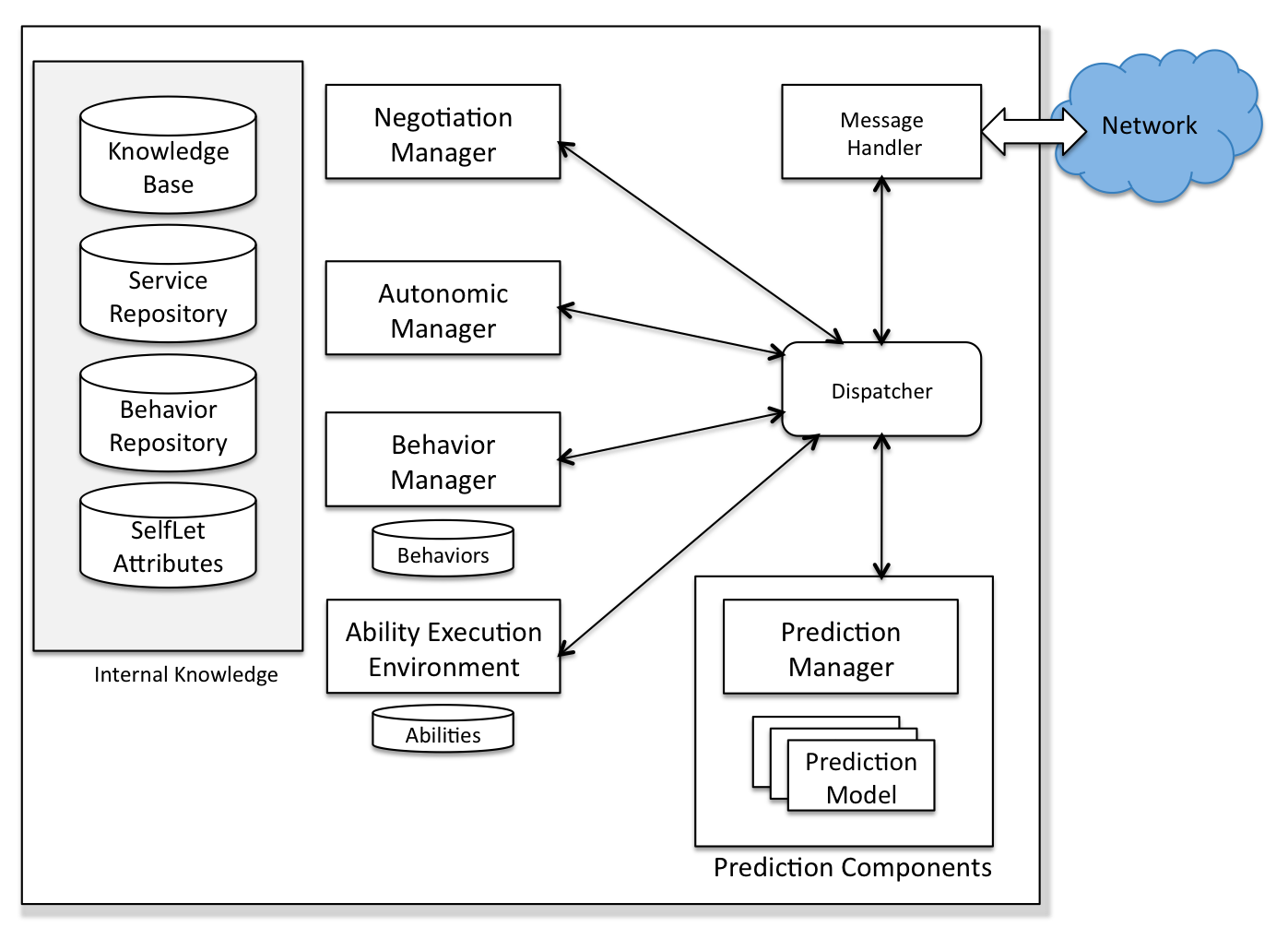} \caption{The internal architecture of a \textit{SelfLet}.}
\label{fig:ACE-architecture}
\end{figure}

The heart of the \textit{SelfLet} architecture (see Figure \ref{fig:ACE-architecture}) is the \textit{Autonomic Manager}, which is responsible for the evolution of a \textit{SelfLet} during its lifetime. The \textit{Autonomic Manager} manages the \textit{SelfLet} behavior according to the set of \textit{Autonomic Policies} it has installed. This set can evolve over time. The \textit{Autonomic Manager} is implemented exploiting \textit{Drools}~\cite{Drools}, a Java \textit{Production Rule System}.  The well-known features of the Production Rule Systems, along  with the advanced nature of the Drools language, allow the \textit{SelfLet} programmers to write sophisticated autonomic policies, using a simple, declarative style. 

The \emph{Behavior Manager} controls the execution of the \textit{SelfLet}'s behavior. The behavior run by the \textit{SelfLet} to achieve its Goal, indeed, contains calls either to local or remote services. In the case a request for a service arises, the Autonomic Manager triggers the execution of a rule that verifies whether the service is locally available or not, and, in this second case, asks the \textit{Negotiation Manager} to retrieve the Service and negotiate with the corresponding \slet{} a proper offer mode. 

The Negotiation Manager communicates with other \textit{SelfLets} in the network by means of the \textit{Message Handler}. The underlying communication framework used to enable the communication among \textit{SelfLets} is REDS~\cite{REDS}, that adopts the publish/subscribe approach. This communication paradigm does not require \textit{SelfLets} to know each other's identity: a relevant advantage in large-scale, self-managing, distributed systems. 

The \textit{Internal Knowledge} is composed of four parts: \textit{Knowledge Base}, which can be used to store and retrieve any kind of information needed by any of the \textit{SelfLet} components; \textit{Service Repository}, which lists Services the \textit{SelfLet} can offer (to itself or to other \textit{SelfLets}); \textit{Behaviors Repository}, which contains all the StateChart the \textit{SelfLet} is able to run; \textit{Attribute Repository}, which stores descriptions about the \textit{SelfLet}.

The \emph{Ability Execution Environment} is in charge of executing simple Java operations, the \emph{Abilities}, than can be activated as part of behaviors for performing specific low-level tasks. 

The communication internal to the \slet{} follows a publish/subscribe approach. All architectural components are connected to a \emph{Dispatcher} that is in charge of receiving subscriptions for events and event publications. It asynchronously deliver all published events to those components that have subscribed to them. The same publish/subscribe paradigm is adopted for the communication among \moreslet{}. In this second case, in order to guarantee scalability, a distributed and optimized dispatching system is adopted~\cite{REDS}.  

The \moreslet{} approach has been implemented both on a full fledged Java platform and on wireless sensors running the TinyOS operating system and adopting the NesC programming language. This second implementation, known as TinySelfLet, implements a smaller set of functionalities due to the physical constraints of the devices and is currently being consolidated~\cite{Panzeri}. 
 
\section{Integrating prediction models into the SelfLet approach}
\label{sec:framework}

As we have already mentioned, we extend the \slet{} architecture by incorporating prediction models that allow the \moreslet{} autonomic policies to be activated in a more timely way.

The context in which \moreslet{} are used is a distributed one with large number of nodes. Due to this requisite, we decided to adopt a completely decentralized approach in which each node can produce predictions and actuate them independently from other nodes.

Any prediction model bases its forecasts on what happened in the past within a certain (sub)system. In our specific case we look at the events that are produced within a \slet{} and at those that arrive to that \slet{} from the external environment. These events are received as input by prediction models that, in turn, generate new events that represent the forecast. Such forecast is then received by the Autonomic Manager and can activate a proper autonomic policy. The framework allows for two degrees of freedom: the first is constituted by the choice of the prediction model to adopt and second one is represented by the definition of the autonomic policy that implements the reaction to a prediction. 

Algorithms implementing new prediction models can be integrated into a \slet{} following a plugin approach. Each implementation should inherit from a specific Java class hierarchy, \texttt{PredictionModel}, and should be accompanied by a descriptor where the events needed for the prediction model to work are specified. The same descriptor includes also the events that are generated by the same implementation and that will be the inputs to trigger some autonomic policy. 
The autonomic policy is defined as a set of Drools production rules and it is installed in the Autonomic Manager according to the mechanisms already defined as part of a \slet{}.

Prediction model implementations are operated by the \emph{Prediction Manager} that is in charge of their correct installation and of providing them with the needed events. Similarly to what happens in the Eclipse framework, at the time the \slet{} is deployed, the Prediction Manager looks for prediction model implementations to install and checks the corresponding descriptors. Then it connects to the \slet{} Dispatcher from which it receives all events being circulated in the \slet{}. When it receives an event of interest of some prediction model implementation, then it forwards such event to it. When a prediction model implementation produces a forecast, it forwards it to the Prediction Manager that, in turn, publishes it to the Dispatcher. If any autonomic policy exists that is triggered by that event, then this receives it and executes the corresponding reaction. The prediction subsystem resulting from the development of Prediction Manager and prediction model implementations is highlighted in Figure \ref{fig:ACE-architecture}.

%

\section{Example and results}
\label{sec:example}

We performed a preliminary validation of our framework with the objective to evaluate that the functionality of the prediction framework is correct. 
The validation of the proposed framework consisted in the following main stages: 1) the development of a simple example involving two types of cooperating \moreslet{}; 2) the individuation of a possible autonomic behavior that could optimize the execution of the \moreslet{} pair; 3) the implementation of the corresponding prediction model and autonomic policy, 4) the execution of the system and the collection of results.  

\subsection{The example}
\begin{figure}[htb]
    \begin{center}
      \includegraphics[width=\columnwidth]{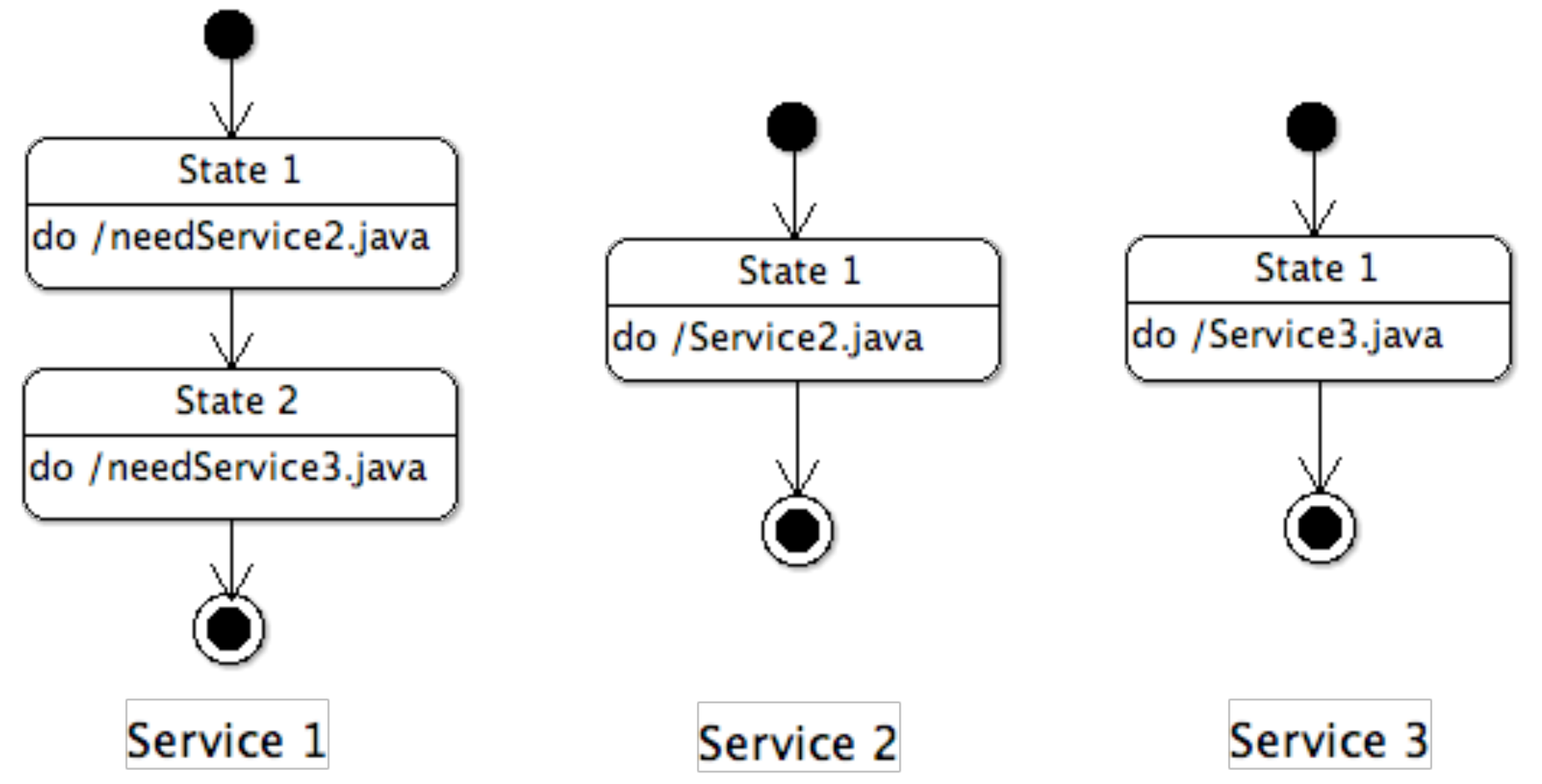}	
      \caption{\label{fig:behaviors} The behaviors installed in $S_2$.}
    \end{center}
  \end{figure}

The example that we have adopted for the evaluation shows a very simple application logic. The reason for this is that we wanted to focus our attention on the prediction and reconfiguration aspects and not on a specific application logic. The system configuration is composed of two \moreslet{}: $S_1$ and $S_2$. $S_1$ needs periodically a service (\texttt{Service 1}) that is offered by $S_2$. In turn, $S_2$ offers three services, \texttt{Service 1}, \texttt{Service 2}, and \texttt{Service 3} that are implemented by the three basic behaviors shown in Figure \ref{fig:behaviors}. Looking at the behaviors, the reader can notice that, in order to execute \texttt{Service 1}, $S_2$ needs to call also \texttt{Service 2} and \texttt{Service 3}. All services are initially offered in the ``Can Do'' mode.

\subsection{Expected autonomic behavior, prediction model, and autonomic policies}
Even though the example described in the previous section is quite simple, there is at least one element that could introduce an optimization in the execution of the system. This is the idea of forcing the learning of frequently requested services so that their future invocations will be directly carried out locally instead of remotely. This allows a saving of network messages and brings computation where it is more likely to be needed.
In our example such optimization could be possibly achieved by having $S_2$ teaching \texttt{Service 1} to $S_1$ with the aim of limiting the number of requests that reach $S_2$ from $S_1$.

In order to achieve this objective, we defined a prediction model that, based on the history of service requests arriving to a certain \slet{} from the other \moreslet{}, can forecast a certain trend of requests and therefore can trigger the execution of an autonomic policy that enables the service to be requested in a ``Can Teach'' mode. The prediction model we have chosen for the experiment is based on the simple assumption that if a certain service request is executed frequently in a certain time window in the immediate past, then it will be very likely reiterated again in the future. Thus, the prediction model implementation requires to know all the events that concern outgoing service requests and produces an event that identifies the service that should be acquired by the \slet{}. 

Many algorithms are available in the literature to identify frequent events in a data stream \cite{CDG08,CM03,CCF02}; we chose to implement the algorithm proposed by Metwally et Al. \cite{MAE06} due to its tight error guarantees and minimal space requirements, desirable property on small devices. To be considered frequent, an event must satisfy the following conditions on the minimum number of events and on the support: 
\begin{equation}
	NumEvents_x \geq  Minimum\:Occurrences  
\end{equation}

\begin{equation}
	Support_x = \frac{NumEvents_x}{NumTotEvents} \geq FrequencyThreshold     
\end{equation}

In our example, such equations are applied to the events representing service requests. The result produced by the prediction model implementation is a new event containing information about the most frequent service request and its support. Such event triggers the policy described in pseudocode in Listing \ref{autonomic_policy}.

\begin{lstlisting}[captionpos=b,language=drools,frame=single,caption=Autonomic policy,label=autonomic_policy ]
rule ``change service ask mode"
	when 
	 	``Requests for a remote service become frequent''
	then
		``Ask the service in teach mode''
end
\end{lstlisting} 

\subsection{Results}
The dynamic behavior we expect to see in the example can be divided in three stages:
\begin{enumerate}
\item The first stage sees $S_1$ asking for service \verb#Service 1#. As requests for such remote service keep increasing, there will be a point in time when the service is considered frequent. As a consequence, the service is asked in \emph{teach} mode and the behavior implementing it becomes also available at $S_1$. However the subservices needed by the just installed behavior are not transferred.

\item The second stage sees \verb#Service 1# locally executed by $S_1$. Since the two subservices needed by the behavior are not located within $S_1$, two broadcast service requests are sent. These will be possibly replied by $S_2$. Now things seems getting worse since to complete the service, two messages need to be sent (i.e. one for \verb#Service 2# and one for \verb#Service 3#). However this is a temporary situation, in fact, if $S_1$ will keep calling the other two services, these eventually will be recognized as frequent services and a request for them in \emph{teach} mode will be issued.

\item In the last step the system converges into a stable situation in which the needed services are all locally available for each node. At this point no more messages, asking it to achieve a remote service, are sent to $S_2$.
\end{enumerate}

The described example has been ran for different time intervals, monitoring the number of times a service has been executed; in order to evaluate the results, the same experiment has been executed also by disabling the self-optimization policy. 
\begin{table}[htdp]
\begin{center}
\begin{tabular}{|c|c|c|c|}
\hline
Experiment	& Goals   	& Goals   	& Total number \\
duration 	& executed 			& executed  & of messages\\
	(sec)		& by $S_1$			& by $S_2$			& exchanged\\\hline

\hline

\hline
\multicolumn{4}{|c|}{Without self-optimizing policy}\\
\hline
180  & 78 & 87 & 78 \\\hline
300  & 132 & 146 &160\\\hline
540  & 240 & 240 & 242\\\hline
900  & 403 & 403 & 403\\\hline
\hline
\multicolumn{4}{|c|}{With self-optimizing policy}\\
\hline

180 & 78 & 87 & 79 \\\hline
300 & 128 & 145 &160\\\hline
540 & 232 & 265 & 311\\\hline
900 & 410 & 444 & 311\\\hline
\end{tabular}
\end{center}
\caption{The results obtained in the example experiment.}
\label{tab:result}
\end{table}
Table \ref{tab:result} shows the results obtained for different time intervals; it reports the cumulative number of times the service has been requested in each \slet{} and the total number of messages exchanged within the system due to the request for remote services.
It is possible to notice that after about 540 seconds, the system with the autonomic policy reaches a stability and no more messages are exchanged between \moreslet{}. Conversely, in the system without the autonomic policy, $S_1$ continues to forward requests to $S_2$ thus having a linear increase of messages with time. It is important to notice that at 540 seconds, the configuration without the autonomic policy sends less messages than the other configuration; this behavior is expected because after having learnt \texttt{Service 1}, $S_1$ needs to send two remote requests for the two subservices. However, this is a transient phase, since the two subservices will be later on recognized as frequent and thus learnt by $S_1$.

\section{Related work}
\label{sec:related_work}

Several research groups are being investigating the various areas of autonomic computing and are building frameworks to support the development of systems showing some self* property. 
Besides the seminal Manifesto ~\cite{manifesto}, IBM contributed to the Autonomic Computing research field with a reference architectural model \cite{blueprint}; this model represents autonomic computing systems as a layered architecture together with an autonomic control loop. A different approach to the autonomic computing is the one known as \emph{emergence}; it takes inspiration from the biological world which contains many examples of successful distributed systems. More precisely, the emergence refers to complex behaviors emerging from the interaction of many elements performing very simple actions. Reference work for this approach are \cite{emergence1} and \cite{emergence2}.
To the best of our knowledge no work explicitly faces the problem of introducing a framework for prediction models into an autonomic platform. 

Referring to more theoretical approaches, an important part of the work performed on prediction models regards the management of Web systems \cite{CAC08,ACC08}. These works study the issues involved in creating a representative view of a web system by detecting significant and non-transient load changes. A similar work has been carried out in \cite{PGC05} in which periods of high utilization or poor performance are predicted using data mining and machine learning techniques. The study aims at optimizing the resources assignment and the computation of opportunistic job scheduling by using auto-regressive methods, multivariate regression methods and bayesian network classifiers. In \cite{KM06} the authors present an approach to obtain response time predictions regarding a web application. However the work does not specify how a component acting as our autonomic manager should use the results of prediction which is left as future work.

\section{Conclusions and lesson learnt}
\label{sec:conclusion}
In this paper we described the extension of an existing autonomic framework with prediction models. We decided to design a flexible architecture in which different prediction models can be hosted. This has been achieved following a plugin approach. In order to validate the work, we implemented a prediction model which computes frequency estimates of service requests issued by \moreslet{}. The reaction to the predictions is described by a policy optimizing the location of more frequently requested services within the system. 
The algorithm that allows a \slet{} to identify frequent service calls is simple on purpose as this way it can be executed on the fly any time it is needed. The proposed autonomic feature has been tested with a simple example; the results showed that after an initial time interval the system converges toward a stable configuration in which frequently executed remote services are taught to the \moreslet{} that invoke them. 
The tests have also highlighted some critical aspect. In particular, the fact that the transfer of a behavior could determine a cascade effect in which many other services are transfered may introduce some dangerous recursion. This aspect could be handled by evaluating in advance the consequences of moving a service to a \slet{}.
How to handle this aspect as well as the identification and incorporation of other prediction models and autonomic policies is the subject of our ongoing research.

\section*{Acknowledgements}
The work presented in this paper is sponsored by the FIRB 2005 project Artdeco and by the European Commission, Programme IDEAS-ERC, Project 227977-SMScom.

\footnotesize
\bibliographystyle{abbrv}
\bibliography{bibliography}  

\begin{thebibliography}{10}

\bibitem{Drools}
http://www.jboss.org/drools/.

\bibitem{automate}
M.~Agarwal, V.~Bhat, H.~Liu, V.~Matossian, V.~Putty, C.~Schmidt, G.~Zhang,
  L.~Zhen, M.~Parashar, B.~Khargharia, and S.~Hariri.
\newblock Automate: Enabling autonomic applications on the grid.
\newblock {\em International Workshop on Active Middleware Services}, 0:48,
  2003.

\bibitem{ACC08}
M.~Andreolini, S.~Casolari, and M.~Colajanni.
\newblock Models and framework for supporting runtime decisions in web-based
  systems.
\newblock {\em ACM Trans. Web}, 2(3):1--43, 2008.

\bibitem{emergence1}
R.~Anthony.
\newblock Emergence: a paradigm for robust and scalable distributed
  applications.
\newblock {\em Proceedings of the 1st International Conference on Autonomic
  Computing}, 2004.

\bibitem{anthill}
O.~Babaoglu, H.~Meling, and A.~Montresor.
\newblock Anthill: A framework for the development of agent-based peer-to-peer
  systems.
\newblock pages 15--22, 2002.

\bibitem{ARAMIS08}
S.~Bindelli, E.~D. Nitto, R.~Mirandola, and R.~Tedesco.
\newblock Building autonomic components: the {SelfLets} approach.
\newblock In {\em Proc. ASE - Workshops}, pages 17--24, 2008.

\bibitem{BJR94}
G.~Box, G.~M. Jenkins, and R.~Gregory.
\newblock {\em Time Series Analysis: Forecasting and Control (3rd edition)}.
\newblock Prentice Hall Engineering, 1994.

\bibitem{CDG08}
T.~Calders, N.~Dexters, and B.~Goethals.
\newblock Mining frequent items in a stream using flexible windows.
\newblock {\em Intell. Data Anal.}, 12(3):293--304, 2008.

\bibitem{CAC08}
S.~Casolari, M.~Andreolini, and M.~Colajanni.
\newblock Runtime prediction models for web-based system resources.
\newblock In {\em Modeling, Analysis and Simulation of Computers and
  Telecommunication Systems, 2008. MASCOTS 2008. IEEE International Symposium
  on}, 2008.

\bibitem{CCF02}
M.~Charikar, K.~Chen, and M.~Farach-Colton.
\newblock Finding frequent items in data streams.
\newblock In {\em ICALP '02: Proceedings of the 29th International Colloquium
  on Automata, Languages and Programming}, pages 693--703, London, UK, 2002.
  Springer-Verlag.

\bibitem{CM03}
G.~Cormode and S.~Muthukrishnan.
\newblock What's hot and what's not: tracking most frequent items dynamically.
\newblock In {\em PODS '03: Proceedings of the twenty-second ACM
  SIGMOD-SIGACT-SIGART symposium on Principles of database systems}, pages
  296--306, New York, NY, USA, 2003. ACM.

\bibitem{REDS}
G.~Cugola and G.~P. Picco.
\newblock Reds: A reconfigurable dispatching system.
\newblock In {\em In Proc. of the 6th Int. Workshop on Software Engineering and
  Middleware SEM06}, pages 9--16. ACM Press, 2006.

\bibitem{ASE07}
D.~Devescovi, , E.~{Di Nitto}, and R.~Mirandola.
\newblock An infrastructure for autonomic system development: the selflet
  approach.
\newblock In {\em Proc. ASE}, pages 449--452, 2007.

\bibitem{FGGPLS}
N.~Friedman, D.~Geiger, M.~Goldszmidt, G.~Provan, P.~Langley, and P.~Smyth.
\newblock Bayesian network classifiers.
\newblock In {\em Machine Learning}, pages 131--163, 1997.

\bibitem{ganek_corbi}
A.~G. Ganek and T.~A. Corbi.
\newblock The dawning of the autonomic computing era.
\newblock {\em IBM System Journal}, 42(1):5--18, 2003.

\bibitem{manifesto}
IBM.
\newblock Autonomic computing manifesto.

\bibitem{blueprint}
IBM.
\newblock An architectural blueprint for autonomic computing.
\newblock 2006 (Fourth Edition).

\bibitem{autonomic-vision}
J.~Kephart and D.~Chess.
\newblock The vision of autonomic computing.
\newblock {\em Computer}, 36:41--50, January 2003.

\bibitem{KM06}
S.~Kirtane and J.~Martin.
\newblock Application performance prediction in autonomic systems.
\newblock In {\em Proceedings of the 44th annual Southeast regional
  conference}, pages 566--572, New York, NY, USA, 2006. ACM.

\bibitem{emergence2}
\mbox{T. De Wolf and T. Holvoet}.
\newblock Emergence as a general architecture for distributed autonomic
  computing.
\newblock 2004.

\bibitem{MAE06}
A.~Metwally, D.~Agrawal, and A.~E. Abbadi.
\newblock An integrated efficient solution for computing frequent and top-k
  elements in data streams.
\newblock {\em ACM Trans. Database Syst.}, 31(3):1095--1133, 2006.

\bibitem{Panzeri}
M.~Panzeri.
\newblock Studio di un approccio per la realizzazione di agenti autonomici in
  reti di sensori wireless.
\newblock Master's thesis, Politecnico Di Milano, 2008.

\bibitem{PGC05}
R.~Powers, M.~Goldszmidt, and I.~Cohen.
\newblock Short term performance forecasting in enterprise systems.
\newblock In {\em KDD '05: Proceedings of the eleventh ACM SIGKDD international
  conference on Knowledge discovery in data mining}, pages 801--807, New York,
  NY, USA, 2005. ACM.

\bibitem{SR06}
D.~Stutzbach and R.~Rejaie.
\newblock Understanding churn in peer-to-peer networks.
\newblock In {\em IMC '06: Proceedings of the 6th ACM SIGCOMM conference on
  Internet measurement}, pages 189--202, New York, NY, USA, 2006. ACM.

\bibitem{VAHMW02}
R.~Vilalta, C.~V.~Apte, J.~L.~Hellerstein, S.~Ma, and S.~M.~Weiss.
\newblock Predictive algorithm in the management of computer systems.
\newblock {\em IBM System Journal}, 41(3):461--474, 2002.

\end{thebibliography}


\begin{thebibliography}{1}

\bibitem{Drools}
http://www.jboss.org/drools/.

\bibitem{ARAMIS08}
S.~Bindelli, E.~D. Nitto, R.~Mirandola, and R.~Tedesco.
\newblock Building autonomic components: the {SelfLets} approach.
\newblock In {\em Proc. ASE - Workshops}, pages 17--24, 2008.

\bibitem{selflet}
D.~D. D.~Devescovi, E. Di~Nitto and R.~Mirandola.
\newblock Self-organization algorithms for autonomic systems in the selflet
  approach.
\newblock {\em Autonomics}, 2007.

\bibitem{ganek_corbi}
A.~Ganek and T.~Corbi.
\newblock The dawning of the autonomic computing era.
\newblock {\em IBM System Journal}, 42(1), 2003.

\bibitem{manifesto}
IBM.
\newblock Autonomic computing manifesto.

\bibitem{autonomic-vision}
J.~Kephart and D.~Chess.
\newblock The vision of autonomic computing.
\newblock {\em Computer}, 36:41--50, January 2003.

\end{thebibliography}
\balancecolumns

\end{document}